\documentclass{article}
\usepackage{umr-V2}
\usepackage[T1]{fontenc}
\usepackage{epic,eepic,amsmath,latexsym,amssymb,color}
\usepackage{ifthen,graphics,epsfig}
\usepackage[english]{babel}
\usepackage{times}
\usepackage{float}
\usepackage{xspace}

\RRNo{2028}

\RRdate{December 2015}

\RRauthor{
Achour Most\'efaoui\thanks{LINA, 
Universit\'e de Nantes, 44322 Nantes, France}
Matoula Petrolia\thanks{LINA, 
Universit\'e de Nantes, 44322 Nantes, France}
Michel Raynal\thanks{Institut Universitaire de France and ASAP  
(\'equipe commune avec l'Universit\'e de Rennes 1 et Inria)}
Claude Jard\thanks{LINA, 
Universit\'e de Nantes, 44322 Nantes, France}
}

\authorhead{A. Most\'efaoui, M. Petrolia, M. Raynal \& Cl. jard}

\RRtitle{M\'emoire partag\'ee fiable 
dans les syst\`emes asynchones avec processus Byzantins}
\RRetitle{ Atomic Read/Write Memory in Signature-free \\
            Byzantine Asynchronous Message-passing Systems}

\RRresume{
 Cet article pr\'esente une construction de  m\'emoire partag\'ee 
 au dessus d'un  syst\`eme asynchone \`a passage de messages dans lequel 
jusqu'\`a $t<n/3$ processus peuvent avoir des comportements arbitraires 
(fautes Byzantines), $n$ \'etant le nombre total de processsus. 
}
\RRabstract{
This article presents a signature-free distributed algorithm  which builds 
an atomic read/write shared memory on top of an $n$-process asynchronous 
message-passing system in which up to $t<n/3$ processes may commit Byzantine 
failures. 
From a conceptual point of view, this algorithm is designed to be as close as 
possible to the algorithm proposed by Attiya, Bar-Noy and Dolev
(JACM 1995), which builds an atomic register in an $n$-process
asynchronous message-passing 
system where up to $t<n/2$ processes may crash. The proposed algorithm is 
particularly simple. It does not use cryptography to cope with Byzantine 
processes, and is optimal from a $t$-resilience point of view 
($t<n/3$). A read operation requires $O(n)$ messages, and a write operation 
requires $O(n^2)$ messages.
}
\RRmotcle{
syst\`emes asynchrones \`a passage de messages, 
registres atomiques read/write, 
diffusion fiable, 
processus Byzantins.
}
\RRkeyword{
Asynchronous  message-passing system, Atomic read/write register, 
Byzantine process, Linearizability, Reliable broadcast abstraction. 
}

\newlength {\squarewidth}

\newtheorem{theorem}{Theorem}
\newtheorem{lemma}{Lemma}

\newcommand{\toto}{xxx}
\newenvironment{proofT}{\noindent{\bf Proof }}
{\hspace*{\fill}$\Box_{Theorem~\ref{\toto}}$\par\vspace{3mm}}
\newenvironment{proofL}{\noindent{\bf Proof }}
{\hspace*{\fill}$\Box_{Lemma~\ref{\toto}}$\par\vspace{3mm}}

\newenvironment{lemma-repeat}[1]{\begin{trivlist}
\item[\hspace{\labelsep}{\bf\noindent Lemma~\ref{#1} }]}%
{\end{trivlist}}

\newenvironment{theorem-repeat}[1]{\begin{trivlist}
\item[\hspace{\labelsep}{\bf\noindent Theorem~\ref{#1} }]}%
{\end{trivlist}}

\newcounter{linecounter}
\newcommand{\linenumbering}{\ifthenelse{\value{linecounter}<10}
{(\arabic{linecounter})}{(\arabic{linecounter})}}
\renewcommand{\line}[1]{\refstepcounter{linecounter}\label{#1}\linenumbering}
\newcommand{\resetline}[1]{\setcounter{linecounter}{0}#1}
\renewcommand{\thelinecounter}{\ifnum \value{linecounter} > 
9 \else \fi\arabic{linecounter}}
\newfloat{algorithm}{thp}{lop}
\floatname{algorithm}{Algorithm}
\newfloat{construction}{thp}{lop}
\floatname{construction}{Construction}
\newcommand{\Xomit}[1]{}
\newcommand{\REG}{\mathit{REG}}

\begin{document}
\makeRR

\section{Introduction}
\paragraph{Shared memory abstraction on top of a message-passing system} 
{\it Informatics} is a science of abstractions, and accordingly 
(as in sequential computing) the writing of distributed applications 
can be greatly facilitated by the design and the use of underlying 
appropriate abstractions.

This paper considers  the design of such an abstraction, namely an atomic
read/write memory, on top of an asynchronous message-passing distributed 
system made up of $n$ processes, and where up to $t$ processes may commit 
failures. 
The case of crash failures was solved 
by Attiya, Bar-Noy and Dolev in~\cite{ABD95} (a) where it is shown that 
$t<n/2$ is an upper bound for the model parameter $t$, and (b) where a simple, 
elegant, and  $t$-resilient  optimal algorithm is proposed. 
This algorithm is called ABD in the following.

This paper focuses on the case where processes may commit Byzantine failures, 
i.e., may behave in a way that does not respect their intended behavior
(as defined by their specification).

\paragraph{Related work}
Considering the {\it clients/servers} distributed model, several
articles have addressed the design of servers implementing a shared
memory accessible by clients. The servers are usually managing a set
of disks (e.g., \cite{CM05,ICKM06,MA04}).  Moreover, while they
consider that some servers can be Byzantine, some articles restrict
the failure type allowed to clients.  As an example,
\cite{DGMSV11,GV06} explore the efficiency issues (relation between
resilience and fast reads) in the context where only servers can be
Byzantine, while clients (the single writer and the readers) can fail
by crashing.  As other examples, \cite{ICKM06}~considers that clients
can only commit crash failures, while \cite{AB06} considers that
clients can only be ``semi-Byzantine'' (i.e., they can issue a bounded
number of faulty writes, but otherwise respect their code). The
algorithm presented in~\cite{MR98} allows clients and some number of
servers to be Byzantine, but requires clients to sign their messages.
As far as we know, \cite{AAB07} was the first paper considering
Byzantine readers while still offering maximal resilience (with respect 
to the number of  Byzantine servers) without using cryptography. 
However, the writer can fail only by crashing, and the fact that a --possibly 
Byzantine-- reader does not write a fake value in a register (to ensure the 
``reader have to write'' rule required to implement atomicity)  
is insured only with some probability.\\

In the {\it peer-to-peer} model (defined here as a model in which all 
processes are ``equal''),  the construction of an atomic register 
requires that each process manages a copy the register 
that is built. 
The first algorithm building a read/write shared memory
in a message-passing system where processes may commit Byzantine 
failures is (to our knowledge) the one presented in~\cite{IRRS14}. 
This paper considers the implementation of an SWMR (single-writer/multi-reader) 
atomic register. It also shows that $t<n/3$ is an upper bound 
the resilience parameter $t$ for such a construction.
In this algorithm,  each SWMR atomic read/write register is represented, 
at each process, by the full history of all its modifications. 

The fact that  an SWMR register is considered is due to the following 
observation: as a Byzantine process can corrupt any register it can write, 
the design of a  multi-writer/multi-reader register with non-trivial
correctness guarantees is impossible in the presence of Byzantine processes.

\paragraph{Content of the paper}
This paper presents a new  algorithm implementing an array of 
$n$ SWMR atomic read/write registers (one per process)
in an asynchronous  message-passing system where up to $t<n/3$ 
processes may commit Byzantine failures.  This algorithm does not require 
to enrich the underlying system with cryptography-based techniques.

When designing this algorithm, an aim was to obtain an algorithm
whose ``spirit'' is ''as close as possible'' to ABD. 
We think that this is important from  both
understanding and pedagogical point of views. It helps better understand 
the ``gap'' between crash failures and Byzantine failures. 
From an algorithmic  point of view, we have the following:
\begin{itemize}
\vspace{-0.2cm}
\item  With respect to the  algorithm described in~\cite{IRRS14}, 
the proposed algorithm requires a process to store only a single pair 
(value, sequence number) per atomic register. 
\vspace{-0.2cm}
\item With respect to ABD, there are two  main differences:
\begin{itemize}
\vspace{-0.2cm}
\item One is the way processes implement the ``reads have to write'' 
requirement needed to obtain the atomicity property of a register~\cite{L86-2}.
\vspace{-0.1cm}
\item The other one lies in the broadcast operation 
used to disseminate new values. While a simple unreliable 
broadcast\footnote{This broadcast is a simple send of the same message 
to all processes. If a process crashes during its execution, it is possible 
that only a subset of the processes receive the message.} 
is sufficient in the presence of process crash failures, a stronger broadcast 
needs to be used to cope with Byzantine processes in a signature-free system. 
\end{itemize}
\end{itemize}
The resulting algorithm is particularly simple. 
Moreover, when considering the non-faulty processes, a read 
costs $O(n)$ messages and  a write  costs $O(n^2)$ messages.

\paragraph{Roadmap}
The paper is composed of~\ref{sec:conclusion} sections. 
Section~\ref{sec:model} presents the computation model, and 
the underlying reliable broadcast abstraction. 
Section~\ref{sec:specification-SWMR} presents a specification of an
SWMR read/write atomic register in the presence of Byzantine processes. 
Then, Section~\ref{sec:construction-RW-register}
presents the algorithm, and Section~\ref{sec:proof} proves its correctness. 
Finally, Section \ref{sec:conclusion} concludes the paper.     

\section{Computation model} 
\label{sec:model}
\subsection{Process model, communication model, and failure model}

\paragraph{Computing entities}
The system is made up of a set $\Pi$ of $n$ sequential processes, 
denoted $p_1$, $p_2$, ..., $p_n$. These processes are asynchronous 
in the sense that each process  progresses at  its own speed, 
which can  be arbitrary and remains always unknown to the other processes. 

\paragraph{Communication model}
The processes cooperate by sending and receiving messages through 
bi-directional channels. The communication network is a complete network, 
which means that each process $p_i$  can directly send a message to any 
process $p_j$ (including itself). It is assumed that the Byzantine processes
cannot control the network, hence when a process receives a  message, 
it can unambiguously identify its sender. 
Each channel is reliable (no loss, corruption, or creation of messages),
not necessarily first-in/first-out, and asynchronous (while the transit 
time of each message is finite,  there is no upper bound bound on message 
transit times).  

A process $p_i$ invokes the operation 
``${\sf send}$ {\sc tag}($m$) {\sf to} $p_j$'' to send the message 
tagged {\sc tag} and carrying the value $m$. It receives a message 
tagged {\sc tag} by invoking the operation ``${\sf receive}$ {\sc tag}()''. 
``${\sf broadcast}$ {\sc tag}($m$)'' is a 
macro-operation that expands  as 
``{\bf for each}  $j\in \{1,\cdots,n\}$  
${\sf send}$ {\sc tag}($m$) {\sf to} $p_j$ {\bf end for}''. 
(The sending  order is arbitrary, which means that, if the sender 
crashes while executing this statement, an arbitrary subset of processes 
of processes will receive the message.)

\paragraph{Byzantine failures}
The model parameter $t$ is an upper bound on the number of
processes that can exhibit a  Byzantine behavior~\cite{LSP82,PSL80}. 
A Byzantine process is a process that behaves
arbitrarily: it can crash, fail to send or receive messages, send
arbitrary messages, start in an arbitrary state, perform arbitrary state
transitions, etc. Hence, a Byzantine process, which is assumed to send a
message $m$ to all the processes, can send a message $m_1$ to some processes, 
a different message $m_2$  to another subset of processes, and no message 
at all to the other processes. Moreover, while they cannot modify the content 
of the messages sent by  non-Byzantine processes, they can read their 
content and reorder their deliveries. More generally,  Byzantine processes 
can collude to ``pollute'' the computation.

A Byzantine process is also called a {\it faulty} process. 
A process that  commits no failure (i.e., a non-Byzantine process) 
is also called a {\it correct} process. 

\paragraph{Notation}
In the following, the previous computation model, restricted to 
the case where $t<n/3$, is denoted ${\cal BAMP}_{n,t}[t<n/3]$. 

\subsection{Reliable broadcast abstraction}
\label{sec:reliable-broadcast}

This section presents a reliable broadcast abstraction 
(denoted r-broadcast) that will be used to build  a read/write register 
(Section~\ref{sec:construction-RW-register}). 
This abstraction is a simple generalization of a reliable broadcast due to 
Bracha~\cite{B87}. While Bracha's abstraction is for a single broadcast, 
the proposed  abstraction considers that each process can issue a sequence 
of broadcasts. It is shown in~\cite{B87} that $t<n/3$ is a necessary 
requirement to cope with the net effect of asynchrony and Byzantine failures.

\paragraph{Specification}
The reliable broadcast abstraction is defined by two operations 
denoted ${\sf R\_broadcast}()$ and ${\sf R\_deliver}()$. 
When a process $p_i$ invokes ${\sf R\_broadcast}()$ we say that  
``$p_i$ r-broadcasts a value''. Similarly, when $p_i$ returns from  
an invocation of ${\sf R\_deliver}()$  and obtains a value, we say 
``$p_i$  r-delivers a value''. 

The operation ${\sf R\_broadcast}()$ has two input parameters: 
a broadcast value $v$, and an integer $sn$, which is a local sequence number
used to identify the successive r-broadcasts issued by each process $p_i$. 
The sequence of numbers  used by each (correct) process is 
the increasing sequence of consecutive integers. 

\begin{itemize}
\vspace{-0.2cm}
\item RB-Validity. If a  correct process r-delivers a  pair $(v,sn)$ from a 
correct process $p_i$, then $p_i$ invoked the operation ${\sf R\_broadcast}(v,sn)$. 
\vspace{-0.2cm}
\item
RB-Integrity. Given any process $p_i$, a correct process r-delivers at 
most once a pair $(-,sn)$  from $p_i$.
\vspace{-0.2cm}
\item RB-Uniformity. If a correct process r-delivers a pair $(v,sn)$ from 
$p_i$ (possibly faulty), then all the correct processes eventually r-deliver 
the same pair $(v,sn)$  from $p_i$. 
\vspace{-0.2cm}
\item RB-Termination. If the process that invokes ${\sf R\_broadcast}(v,sn)$ 
is correct, all the correct processes eventually r-deliver the pair $(v,sn)$.
\end{itemize}
RB-Validity is on correct processes and relates their outputs to their 
inputs, namely no correct process r-delivers spurious messages from 
correct processes. 
RB-Integrity states  that there is no r-broadcast duplication. 
RB-Uniformity is an ``all or none'' property (it is not possible for a pair to 
be delivered by a correct process and to be never delivered by the other
correct  processes).  RB-Termination is a liveness property: at least all 
the pairs r-broadcast by correct processes are r-delivered by them. 

For completeness,  an algorithm (due to Bracha~\cite{B87}), which  implements
the r-broadcast abstraction in the model  ${\cal BAMP}_{n,t}[t<n/3]$, 
is described in Appendix~\ref{sec:algorithm-RB}.


\section{Atomic Read/Write Registers in the Presence of Byzantine Processes}
\label{sec:specification-SWMR}

\subsection{Definitions and specification}
\label{sec:specification}
\paragraph{Single-writer/multi-reader (SWMR) registers} 
The fault-tolerant shared memory supplied to the upper abstraction layer 
is an array denoted $\REG[1..n]$. 
For each $i$,  $\REG[i]$ is a single-writer/multi-reader (SWMR) register. 
This means that  $\REG[i]$ can be written only by $p_i$. To that end, 
$p_i$  invokes  the operation $\REG[i].{\sf write}(v)$ 
where $v$ is the value it wants to  write into   $\REG[i]$. Differently, 
any process $p_j$  can read  $\REG[i]$. It invokes then the operation 
$\REG[i].{\sf read}()$.

As already noticed in the Introduction, the ``single-writer'' requirement is  
natural in the presence of Byzantine processes. If registers could be written 
by any process, it would be possible for the Byzantine processes to 
pollute the whole memory, and no non-trivial computation could be possible.

\paragraph{On write operations by Byzantine processes} 
A Byzantine process $p_k$ may invoke the write operation 
$\REG[k].{\sf write}()$ 
to assign a  new value to $\REG[k]$, but it may also try to modify 
$\REG[k]$  without using this operation. In such a case, its fraudulent
attempt to modify $\REG[k]$  may succeed or not. If it succeeds, 
the corresponding modification of  $\REG[k]$  is considered 
(from an external observer's point of view) as if it  has been produced by an 
invocation of $\REG[k].{\sf write}()$\footnote{As we will see, 
at the operational level, when a modification of $\REG[i]$ by a Byzantine 
process $p_i$ succeeds, the underlying messages generated by $p_i$ could 
have been sent by a correct implementation of the operation ${\sf write}()$.}. 
This is because no correct process  can distinguish such a modification of 
$\REG[k]$ from a  call to the write  operation  by $p_k$. Let us nevertheless
notice that this does not prevent the  fact that the value assigned to  
$\REG[k]$ can be a fake value. Moreover,  at the abstraction level defined 
by $\REG[1..n]$, as  $p_k$ is sequential,  its modifications of $\REG[k]$ 
appear as if they have been executed sequentially.

\paragraph{Definitions}
\begin{itemize}
\vspace{-0.2cm}
\item
A sequence of values, denoted $H_i$, is associated with each register $\REG[i]$.
$H_i$ is the sequence of values written in $\REG[i]$.
Let $H_i[x]$ denote the $x^{\mbox{\small th}}$ element of $H_i$. 
\vspace{-0.2cm}
\item 
The following notations are used. 
\begin{itemize}
\vspace{-0.2cm}
\item Let $p_i$ be a correct process. $read[i,j,x]$:
 execution of $\REG[j].{\sf read}()$ returning  $H_i[x]$. 
\vspace{-0.1cm}
\item $write[i,x]$: 
$x^{\mbox{\small th}}$ update  of $\REG[i]$ by $p_i$. 
Hence, $write[i,x]$ defines the value of  $H_i[x]$.

If $p_i$ is a correct process,  $write[i,x]$ corresponds to an execution of 
 $\REG[i].{\sf write}()$. 
If $p_i$ is Byzantine, according to the previous discussion,
these  ``$write[i,x]$''  capture all the modifications of 
$\REG[i]$ by $p_i$, be them associated with a call to the write operation 
or not. (Let us remember that, at this abstraction level, any process is 
sequential.)
\end{itemize}
\end{itemize}

\paragraph{Specification}~\\
The correct behavior of the array of registers $\REG[1..n]$ is defined 
by the following set of properties. 
\begin{itemize}
\vspace{-0.2cm}
\item Termination (liveness). Let $p_i$ be a correct process. 
\begin{itemize}
\vspace{-0.2cm}
\item Each invocation of  $\REG[i].{\sf write}()$ terminates. 
\vspace{-0.1cm}
\item  For any $j$,  any invocation of $\REG[j].{\sf read}()$ by 
$p_i$ terminates. 
\end{itemize}
\item Consistency (safety)\footnote{It would be possible to associate 
a start event and an end event with each $read[i,j,x]$ and each $write[i,y]$
issued by a correct process, and a start event with each $write[j,y]$
issued by a Byzantine process, 
so that all these events define a total order from which  the  notion of  
``terminates before''  could be formally defined (as in~\cite{HW90,L86,R13}). 
To not overload the presentation, we do not use this formalization here.}. 
Let $p_i$ and $p_j$ be  correct processes and $p_k$ any process.
\begin{itemize}
\vspace{-0.2cm}
\item Read followed by write: 
($read[i,k,x]$ terminates before  $write[k,y]$ starts) $\Rightarrow$ $(x<y)$. 
\vspace{-0.1cm}
\item Write followed by read: 
($write[j,x]$ terminates before  $read[i,j,y]$ starts) 
$\Rightarrow$ $(x\leq y)$.\footnote{Let us notice that this property considers 
that the write of $\REG[j]$ is issued by a correct process. This is because 
it is not always possible to define when the modification of $\REG[j]$ has 
terminated when $p_j$ is Byzantine.}
\vspace{-0.1cm}
\item No read inversion: 
($read[i,k,x]$ terminates before  $read[j,k,y]$ starts) 
$\Rightarrow$ $(x\leq y)$. 
\end{itemize}
\end{itemize}

As there is no way to constrain the behavior of a Byzantine process, 
the termination property is only on correct processes, and there is no 
requirement on the value returned by a read issued by a Byzantine process. 
The safety property  concerns only the values read by correct processes. 
The first property states that there is no read from the future, while
the second property states that  no read can obtain an overwritten value. 
Due to the possible concurrent accesses to a same register, these  
two properties actually defines a regular register~\cite{L86}. 
Hence the ``no read inversion'' property, which  allows to obtain an
atomic register  from a regular register~\cite{CKW00,L86,R13}. 

\subsection{Linearizability}
Atomic registers were formally defined in~\cite{L86,M86}. 
Then, the  atomicity concept  was extended to any concurrent 
object defined by a sequential specification under the name 
linearizability~\cite{HW90}. Hence in our context, the terms 
``atomic register'' and ``linearizable register'' are synonymous.
The properties provided by linearizability are investigated in~\cite{HW90}.

\paragraph{Definition}
Given a register $\REG[i]$,  {\it linearizability}~\cite{HW90} 
means that it is  possible to totally order the executions of its
read and write operations in such a way that (a) each operation  
appears as if it has been executed at a single point of the time line between 
its start event and its end event, (b)  no two operation executions appear 
at the  same point, and (c) each read operation returns the value written 
by the closest  write operation that precedes it in the sequence
(or the initial value if there is no such write operation).

A register is {\it linearizable} if its operations satisfy the previous 
items (a), (b), and (c).
The  {\it linearization point} of an operation is the point of the timeline
at which this operation appears to have been instantaneously executed.

\paragraph{An important property of Linearizability}
An important theorem associated with linearizability is the 
following~\cite{HW90}: If each object (here a register) is linearizable, then 
the set of all the objects, considered as a single object, is linearizable. 
This means that linearizable objects compose for free.

\begin{theorem}
\label{theorem:linearizable}
The register specification defined in Section~{\em\ref{sec:specification-SWMR}}
defines atomic (linearizable) registers. 
\end{theorem}

\begin{proofT}
As linearizable (atomic) objects compose for free~\cite{HW90}, it is 
sufficient to consider a single register and, starting  from its specification 
defined in Section~\ref{sec:specification},  show that it is linearizable. 

Let $\REG[i]$ be a register. Let $H_i$ be the sequence of values 
written by $p_i$ in  $\REG[i]$.\footnote{As we have seen, if $p_i$ is 
Byzantine, this sequence contains all  the modifications of $\REG[i]$ 
which cannot be distinguished by the correct processes from invocations 
of $\REG[i].{\sf write}()$ by $p_i$.}  
%
The proof consists in building a sequence $S_i$ which 
(a) includes all the read operations of $\REG[i]$ issued  by the correct  
processes plus the writes of $\REG[i]$ issued by $p_i$, and 
(b)  satisfies  the definition of linearizability.

To simplify and without loss of generality, let us assume that 
there is an initial write that gives $\REG[i]$ its initial value. 
Let us start with  $S_i$ being  the sequence of write operations
that produced the sequence $H_i$. 

 Let $read[j,i,x]$ be a read operation 
issued by a correct process $p_j$. 
Let  $write[i,a]$ be the last  write of $\REG[i]$ 
that terminates before  $read[j,i,x]$ starts. 
Let  $write[i,a+1]$, ..., $write[i,a+c]$ 
be (if any) the writes of $\REG[i]$ that are concurrent with  $read[j,i,x]$.
If there is no such writes $c=0$. 
Let $b=c+1$. Hence, if any, $write[i,b]$ is the first write of $\REG[i]$ 
starting after  $read[j,i,x]$ has terminated. We have the following. 
\begin{itemize}
\vspace{-0.2cm}
\item  It follows that from the properties 
``read followed by write'' and  ``write followed by read'' that 
$x\in \{a, a+1, ..., a+c\}$.
\vspace{-0.2cm}
\item 
It follows from  the ``no read inversion'' property that 
if  $read[\ell ,i,x']$  (issued a correct process $p_\ell$)  starts
after $read[j,i,x]$, we have $x\leq x'$. 
\end{itemize}
The operation  $read[j,i,x]$ is added to $S_i$ just after $write[i,x]$. 
It there are two (or more) operations  $read[j_1,i,x]$  and  
$read[j_2,i,x]$ issued by correct processes, they are placed one after 
the other in the sequence $S_i$. 
All the read operations issued by the correct processes are added to $S_i$
as described. 

It is easy to see that the execution associated with $S_i$ is 
linearizable (i.e., satisfies the items (a), (b), and (c) stated above). 
\renewcommand{\toto}{theorem:linearizable}
\end{proofT}

\section{Construction of  Single-Writer/Multi-Reader Atomic  Registers}
\label{sec:construction-RW-register}
An algorithm constructing an SWMR atomic (linearizable) register
in the presence of up to $t$ Byzantine processes,  
is described in Figure~\ref{algo:RB-SN-based-SWMR-atomic-register}. 
As it assumes $t<n/3$, this algorithm is suited for the 
computing model ${\cal BAMP}_{n,t}[t<n/3]$. The algorithm presents the 
code associated with a correct process $p_i$. 

The design of the algorithm strives to be as close as possible to the 
ABD algorithm~\cite{ABD95}, which implements an atomic register 
in an asynchronous system where at most $t<n/2$ may crash.\footnote{In 
addition to the stronger necessary and sufficient condition $t<n/3$, this 
presentation style allows people aware of the ABD algorithm  to see the
additional statements needed to go from crash failures to Byzantine behavior.}
It uses a {\bf wait}$(condition)$ statement. The corresponding 
process is blocked until the predicate $condition$ becomes satisfied. 
While a process is blocked, it can process the messages it receives.

\paragraph{Local variables}
Each process $p_i$ manages the following local variables whose scope 
is the full computation (local variables are denoted with lower case letters, 
and sub-scripted by the process index $i$).   
\begin{itemize}
\vspace{-0.2cm}
\item $reg_i[1..n]$ is the local representation of the array $\REG[1..n]$ of
atomic SWMR registers. Each local register $reg_i[j]$ 
contains two fields, a sequence number $reg_i[j].sn$, and the 
corresponding value  $reg_i[j].val$. It is initialized to the pair 
$\langle  init_j,0 \rangle$, where $init_j$ is the initial value of 
$\REG[j]$. 
\vspace{-0.2cm}
\item $wsn_i$ is an integer, initialized to $0$, used by $p_i$ to associate 
sequence numbers with its successive write invocations.  
\vspace{-0.2cm}
\item $rsn_i[1..n]$ is an array of sequence numbers (initialized to 
$[0,\cdots,0]$) such that
$sn_i[j]$ is used by $p_i$ to identify its successive read invocations
of $\REG[j]$.\footnote{If we assume that no correct process $p_i$ reads its 
own register  $\REG[i]$, $rsn_i[i]$ can be used to store $wsn_i$.}
\end{itemize}

\paragraph{The operation  $\REG[i].{\sf write}(v)$}
This operation is implemented by the client 
lines~\ref{RB-SWMR-H01}-\ref{RB-SWMR-H04}
and the server lines~\ref{RB-SWMR-H11}-\ref{RB-SWMR-H12}
(which are similar to the algorithm implementing a write operation 
in a crash-prone system~\cite{ABD95}). 

Process $p_i$ first increases $wsn_i$ and r-broadcasts the 
message {\sc write}$(v,wsn_i)$. Let us remark that this is the only 
use of the reliable broadcast abstraction by  the algorithm. 
The process $p_i$  then waits for acknowledgments 
(message {\sc write\_done}$(v,wsn_i)$) from  $(n-t)$ distinct 
processes, and finally terminates the write operation. 
As we will see (Lemma~\ref{lemma:quorum-intersection}), 
the intersection of any two quorums of $(n-t)$ processes 
contains  at least $(t+1)$ correct processes. This intersection property 
will be used to  prove the consistency of the register $\REG[i]$.

When $p_i$  r-delivers a message {\sc write}$(v,wsn)$ from a process $p_j$, 
it waits until  $wsn=reg_i[j]+1$ (line~\ref{RB-SWMR-H12}).  Hence, whatever 
the sender $p_j$, its messages {\sc write}$()$ are processed in their 
sending order.
When this predicate becomes true, $p_i$ updates  accordingly its local 
with respect to  $\REG[j]$ (line~\ref{RB-SWMR-H13}), and sends back 
 to $p_j$ an acknowledgment to inform it that its new write has 
locally  been taken into account (line~\ref{RB-SWMR-H14}).


\begin{figure}[th!]
\centering{\fbox{
\begin{minipage}[t]{150mm}
\footnotesize
\renewcommand{\baselinestretch}{2.5}
\resetline
\begin{tabbing}
aaaaa\=aaa\=aaaaa\=aaaaaa\=\kill

{\bf local variables initialization:}\\
\> $reg_i[1..n]\gets
 [\langle init_0,0\rangle, \dots, \langle init_n,0\rangle]$;
 $wsn_i \gets 0$;  $rsn_i[1..n]\gets [0,\cdots,0]$.\\

\%-------------------------------------------------------------------------------------------------\\~\\

{\bf operation} $REG[i]$.$\mathsf{write}$($v$) {\bf is}\\
\line{RB-SWMR-H01} 
\> $wsn_i\gets wsn_i+1$;\\

\line{RB-SWMR-H02} 
\> $\mathsf{R\_broadcast}$ {\sc write}($v,wsn_i$);\\

\line{RB-SWMR-H03} 
\> {\bf wait} {\sc write\_done}($wsn_i$) 
  $\mathsf{received}$ from  $(n-t)$ different processes;\\
\line{RB-SWMR-H04} 
\>  ${\sf return}()$\\
{\bf end operation}.\\

~\\
{\bf operation} $REG[j]$.$\mathsf{read}$() {\bf is}\\
\line{RB-SWMR-H05} 
\> $rsn_i[j]\gets rsn_i[j]+1$;\\

\line{RB-SWMR-H06} 
\> $\mathsf{broadcast}$ {\sc read}($j,rsn_i[j]$);\\

\line{RB-SWMR-H07} 
\> {\bf wait} 
          \big($reg_i[j].sn\geq  \max(wsn_1, ..., wsn_{n-t})$ 
              where $wsn_1, ..., wsn_{n-t}$ are from  \\

~~~~~~~~~~~~~~~~~~~~~~messages
      {\sc state}$(rsn_i[j],-)$ received from $n-t$ different processes\big);\\

\line{RB-SWMR-H08} 
\>  {\bf let} $\langle w,wsn \rangle$ the value of $reg_i[j]$
    which allows the previous wait to terminate;\\

\line{RB-SWMR-H09}
 \>  $\mathsf{broadcast}$ {\sc catch\_up}($j,wsn$);\\

\line{RB-SWMR-H10}
 \> {\bf wait} \big({\sc catch\_up\_done}($j,wsn$) received from
         $(n-t)$ different processes\big);\\

\line{RB-SWMR-H11} 
\>  ${\sf return}(w)$\\
{\bf end operation}. \\
\%-------------------------------------------------------------------------------------------------\\~\\

{\bf when a message} {\sc write}($v,wsn$)
                {\bf is} $\mathsf{R\_delivered}$  {\bf from} $p_j$ {\bf do}\\

\line{RB-SWMR-H12} 
\>  ${\sf wait} (wsn = reg_i[j].sn+1)$;\\
\line{RB-SWMR-H13} 
\>    $reg_i[j] \leftarrow \langle v,wsn\rangle$;\\


\line{RB-SWMR-H14} 
\> $\mathsf{send}$ {\sc write\_done}($wsn$)  ${\sf to}$ $p_j$.\\

~\\
{\bf when a message} 
{\sc read}$(j,rsn)$ {\bf is} $\mathsf{received}$ {\bf from} $p_k$ {\bf do}\\

\line{RB-SWMR-H15} 
\> $\mathsf{send}$ {\sc state}$(rsn,reg_i[j].sn)$ $\mathsf{to}$ $p_k$.\\

~\\
{\bf when a message} {\sc catch\_up}$(j,wsn)$ 
   {\bf is} $\mathsf{received}$ {\bf from} $p_k$ {\bf do}\\

\line{RB-SWMR-H16} 
\> ${\sf wait}$ $(reg_i[j].sn \geq wsn)$;\\

\line{RB-SWMR-H17} 
\> $\mathsf{send}$ {\sc catch\_up\_done}$(j,wsn)$ $\mathsf{to}$ $p_k$.

\end{tabbing}
\end{minipage}
}
\caption{Atomic SWMR Registers 
         in ${\cal BAMP}_{n,t}[t<n/3]$  (code for process $p_i$)}
\label{algo:RB-SN-based-SWMR-atomic-register}
}
\end{figure}

\paragraph{Write of  $\REG[j]$ by a Byzantine process $p_j$}
Let us observe that the only way for a process $p_i$ to modify $reg_i[j]$ 
is to r-deliver a message {\sc write}$(v,wsn)$ from a (correct or faulty) 
process  $p_j$. 
Due to the RB-Uniformity of the r-broadcast abstraction, it follows that,
if a correct process $p_i$ r-delivers such a message, all correct 
processes will r-deliver the same message, be its sender correct 
or faulty.  Consequently each of them will eventually 
execute the statements of lines~\ref{RB-SWMR-H12}-\ref{RB-SWMR-H14}.

Hence, when a faulty process invokes ${\sf R\_broadcast}$ {\sc write}$(v,wsn)$  
(be the r-broadcast invocation involved in an invocation of  
$\REG[j].{\sf write}(v)$ or not), 
its faulty behavior is restricted to  broadcast fake values for $v$ and $wsn$. 
%

\paragraph{The operation  $\REG[j].{\sf read}()$}
This operation  is implemented  by  the client
lines~\ref{RB-SWMR-H05}-\ref{RB-SWMR-H11} and the server 
line~\ref{RB-SWMR-H15}. 
The corresponding algorithm is the core of 
the implementation of an SWMR atomic register. 

When $p_i$ wants to read $\REG[j]$, it first broadcasts a read request 
(message {\sc read}$(j,rsn_i[j])$), and waits for corresponding 
acknowledgments (message {\sc state}$(rsn_i[j],-)$). Each of these 
acknowledgment  carries the sequence number associated with the current value 
of  $\REG[j]$, as known by the  sender of the message
(line~\ref{RB-SWMR-H15}). 
For $p_i$ to progress, the wait predicate (line~\ref{RB-SWMR-H07}) 
states that  its local representation of $\REG[j]$, namely $reg_i[j]$,
must be fresh enough (let us remember that the only line where 
$reg_i[j]$ can be modified is line~\ref{RB-SWMR-H13}, i.e., when $p_i$
r-delivers a message {\sc write}$(-,-)$ from $p_j$). 
This {\it freshness} predicate states that $p_i$'s current value 
of $reg_i[j]$ is as fresh as the current value of at least $(n-t)$ processes
(i.e., at least $(n-2t)$ correct processes). 
If the freshness predicate is false, it will become true when $p_i$ will have 
r-delivered {\sc write}$(-,-)$ messages, which have been r-delivered by other 
correct processes, but not yet  by it. 

When this waiting period terminates, $p_i$ considers the current value 
$\langle w, wsn\rangle$ of  $reg_i[j]$ (line~\ref{RB-SWMR-H08}). 
It then broadcasts the message {\sc catch\_up}$(j,wsn)$, and returns 
the value $w$ as soon as its message {\sc catch\_up}$()$ is acknowledged 
by $(n-t)$ processes  (lines~\ref{RB-SWMR-H09}-\ref{RB-SWMR-H10}). 

The aim of the {\sc catch\_up}$(j,wsn)$ message is to allow each destination 
process $p_k$ to  have a value in its local representation of $\REG[j]$ 
(namely $reg_k[j].val$) at least as recent as the one whose sequence number 
is $wsn$ (line~\ref{RB-SWMR-H15}). 
The aim of this  {\it value resynchronization} is to prevent read inversions. 
When $p_i$ has received the $(n-t)$ acknowledgments it was waiting for
(line~\ref{RB-SWMR-H10}), it knows that no other correct process can obtain 
a value older than the value $w$ it returns. 

\paragraph{Message cost of the algorithm}
In addition to a reliable broadcast (whose message cost is $O(n^2)$), a write 
operation generates $n$ messages {\sc write\_done}. Hence the cost of a 
write is $O(n^2)$ message. A read operation cost $4n$ messages, $n$ messages for each of the four kinds of messages {\sc read}, {\sc state}, {\sc catch\_up} and ${\sc catch\_up\_done}$.

\paragraph{Comparing with the crash failure model}
It is known that the algorithms implementing an atomic register on top of an 
asynchronous message-passing system prone to process crashes, require that 
``reads have to write''~\cite{A00,ABD95,AW04,L86-2,R10}. 
More precisely, before returning a value, in one way or another, a reader
must  write this value to ensure atomicity (otherwise, we have only 
a ``regular'' register~\cite{L86}).  Doing so, it is not possible 
that two sequential read invocations, concurrent with one or more
write invocations, be such that the first read obtains a value while 
the second read  obtains older value (this prevents {\it read inversion}).  

As Byzantine failures are more severe than crash failures, the algorithm of
Figure~\ref{algo:RB-SN-based-SWMR-atomic-register}  needs to use 
a mechanism  analogous to the ``reads have to write'' to prevent read
inversions  from occurring. As previously indicated, 
This is done by the messages {\sc catch\_up}$()$  broadcast at 
line~\ref{RB-SWMR-H09}
and the associated acknowledgments messages  {\sc catch\_up\_done}()
received at line~\ref{RB-SWMR-H10}. As previously indicated, these
messages realize a  synchronization during which 
$(n-t)$ processes (i.e., at least $(n-2t)$ correct processes) 
have resynchronized their value, if needed (line~\ref{RB-SWMR-H15}).

A comparison of two instances of the ABD algorithm~\cite{ABD95} 
and the algorithm of Figure~\ref{algo:RB-SN-based-SWMR-atomic-register}
is presented in Table~\ref{table:comparison}. 
The first instance is the  version of the ABD algorithm which builds 
an array of $n$  SWMR (single-writer/multi-reader) atomic registers 
(one register  per process). 
The second instance  is the version of the ABD algorithm which builds 
a single MWMR  (multi-writer/multi-reader) atomic register. 

As they depend on the application and not on the algorithm, the size of 
the values which are written is considered as constant. 
The parameters $n$ and $t$ have the same meaning as before;
$m$ denotes an upper bound on the number of read and write operations
on each register.  The value $\log n$ is due to the fact that a message 
carries a constant number of process identities. 
Similarly, $\log m$ is due to the fact that 
(a) a message carries a constant number of sequence numbers, and 
(b) there is a constant number of message tags (including the underlying 
reliable broadcast).

\begin{table}[ht]
\begin{center}
\renewcommand{\baselinestretch}{1}
\small
\begin{tabular}{|l|c|c|c|c|c|c|}
\hline
algorithm   & failure type 
    & requirement & msgs/write & msgs/read & msg size & local mem./proc.\\
\hline 
ABD: $n$ SWMR & crash & $t<n/2$   
                  & $O(n)$   &  $O(n)$ &  $O(\log n +\log m)$ &  $O(n\log m)$\\
\hline 
ABD: 1 MWMR & crash &  $t<n/2$ & $O(n)$ 
                  & $O(n)$   & $O(\log n +\log m)$ &  $O(n\log m)$\\
\hline 
Fig.~\ref{algo:RB-SN-based-SWMR-atomic-register}: $n$ SWMR 
        &  Byzantine &  $t<n/3$ &  $O(n^2)$  
              &  $O(n)$ &  $O(\log n +\log m)$ & $O(n\log m)$\\
\hline 
\end{tabular}
\end{center}
\caption{Crash vs Byzantine failures: cost comparisons} 
\label{table:comparison} 
\end{table}

\section{Proof of the construction}
\label{sec:proof}
The model assumption $n>3t$ is implicit in all the statements and proofs 
that follow.  

\subsection{Preliminary lemmas}
\begin{lemma}
\label{lemma:r-delivery}
If a correct process $p_i$ r-delivers a message {\sc write}$(w,sn)$
(from a correct or faulty process), any correct process  r-delivers it. 
\end{lemma}

\begin{proofL}
This is an immediate consequence of the RB-Uniformity property of the
the r-broadcast abstraction. 
\renewcommand{\toto}{lemma:r-delivery}
\end{proofL} %

\begin{lemma}
\label{lemma:quorum-intersection}
Any two sets (quorums) of $(n-t)$ processes have at least 
on correct process in their intersection. 
\end{lemma}

\begin{proofL}
Let $Q_1$ and $Q_2$ be two  sets of processes such that $|Q_1|=|Q_2|=n-t$. 
In the worst case, the $t$ processes that are not in $Q_1$ belong to $Q_2$, 
and the $t$ processes that are not in $Q_2$ belong to $Q_1$.
It follows that
$|Q_1\cap Q_2|\geq n-2t$. As $n>3t$, it follows that 
$|Q_1\cap Q_2|\geq n-2t\geq t+1$, which concludes the proof of the lemma.
\renewcommand{\toto}{lemma:quorum-intersection}
\end{proofL}

\subsection{Proof of the termination properties}

\begin{lemma}
\label{lemma:write-termination}
Let $p_i$ be a correct process. 
Any invocation of $\REG[i].{\sf write}()$ terminates. 
\end{lemma}

\begin{proofL}
Let us consider the first invocation of  $\REG[i].{\sf write}()$ by a
correct process $p_i$. This write operation generates the  r-broadcast 
of  message {\sc write}$(-,1)$ (lines~\ref{RB-SWMR-H01}-\ref{RB-SWMR-H02}).
Due to Lemma~\ref{lemma:r-delivery}, all correct processes r-deliver 
this message, and the waiting predicate of  line~\ref{RB-SWMR-H13} is
eventually satisfied.  Consequently, 
each correct process $p_k$  eventually sets  $reg_k[i].sn$ to $1$, and
sends back to $p_i$ an acknowledgment message {\sc write\_done}$(1)$. 
As there are least $(n-t)$ correct processes, 
$p_i$ receives such acknowledgments from at least $(n-t)$ 
different processes, and  terminates its first invocation 
(lines~\ref{RB-SWMR-H03}-\ref{RB-SWMR-H04}).

As, for any given any process $p_j$, 
all correct processes process the messages {\sc write}$()$  from $p_j$
in their sequence order, the lemma follows from a simple induction
(whose previous paragraph is the proof of the base case).

\renewcommand{\toto}{lemma:write-termination}
\end{proofL} 

\begin{lemma}
\label{lemma:read-termination}
 Let $p_i$ be a correct process. 
For any  $j$, any invocation of  $\REG[j].{\sf read}()$ terminates. 
\end{lemma}

\begin{proofL}
When a correct process $p_i$ invokes   $\REG[j].{\sf read}()$, 
it broadcasts a message {\sc read}$(j,rsn)$ where $rsn$ is a new 
sequence number (lines~\ref{RB-SWMR-H05}-\ref{RB-SWMR-H06}).
Then, it waits until the freshness predicate of line~\ref{RB-SWMR-H07} 
becomes satisfied. As $p_i$ is correct, each correct process $p_k$ receives 
{\sc read}$(j,rsn)$, and sends back to $p_i$ a message 
{\sc state}$(rsn,wsn)$, where $wsn$ is the sequence number of 
the last value of $\REG[j]$ it knows (line~\ref{RB-SWMR-H15}). 
It follows that $p_i$ receives a message {\sc state}$(j,-)$ from at least 
$(n-t)$ correct processes. Let 
{\sc state}$(j,wsn_1)$, $\cdots$, {\sc state}$(j,wsn_{n-t})$ be these messages.  

To show that the wait of line~\ref{RB-SWMR-H07} terminates 
we have to show that the freshness predicate 
$reg_i[j].sn \geq \max(wsn_1,\cdots,wsn_{n-t})$ is eventually satisfied. 
Let $wsn$ be one of the previous sequence number, and $p_k$ the correct 
process that send it. This means that $reg_k[j].sn=wsn$
(line~\ref{RB-SWMR-H15}), from which we conclude (as $p_k$ is correct)
that $p_k$ has previously r-delivered a message {\sc write}$(-,wsn)$ 
and updated accordingly $reg_k[j]$ at line~\ref{RB-SWMR-H13} 
(let us remember that this is the only line at which the local register
 $reg_k[j]$ is updated). It follows from Lemma~\ref{lemma:r-delivery} 
that eventually $p_i$ r-delivers  the message {\sc write}$(-,sn)$.
It follows then from  line~\ref{RB-SWMR-H13} that eventually 
we have $reg_i[j].sn\geq sn$. As this is true for any sequence number 
in $\{wsn_1,...,wsn_{n-t}\}$, it follows that the freshness predicate 
is eventually satisfied, and consequently  the wait statement of 
line~\ref{RB-SWMR-H07} is satisfied. 

Let us now consider the wait statement of line~\ref{RB-SWMR-H10}, 
which appears after $p_i$ has broadcast the message {\sc catch\_up}$(j,wsn)$, 
where $wsn=reg_i[j].sn$ (sequence number in $reg_i[j]$ just after 
$p_i$ stopped waiting at line~\ref{RB-SWMR-H07}).   
We show that any correct process sends back to $p_i$ an acknowledgment 
{\sc catch\_up\_done}$(j,wsn)$ at line~\ref{RB-SWMR-H17}.    
Process $p_i$ updated  $reg_i[j].sn$ to $wsn$ at line~\ref{RB-SWMR-H13},  
and this occurred when it r-delivered a message {\sc write}$(-,wsn)$.
The reasoning is the same as in the previous paragraph, namely,  
it follows from Lemma~\ref{lemma:r-delivery} that all correct processes
r-deliver  this message and consequently we have  $reg_k[j].sn\geq wsn$ 
at every correct process $p_k$. Hence, the value resynchronization 
predicate of line~\ref{RB-SWMR-H16} is eventually satisfied at 
all correct processes, that consequently sends back a message 
{\sc catch\_up\_done}$(j,wsn)$ at line~\ref{RB-SWMR-H17}, which concludes
the proof of the lemma.
\renewcommand{\toto}{lemma:read-termination}
\end{proofL}

\subsection{Proofs of the consistency  (atomicity) properties}

The next lemma shows that a sequence $H_i$, as defined in 
Section~\ref{sec:specification-SWMR},
can be associated with each register $\REG[i]$. 

\begin{lemma}
\label{lemma:history}
Given any register $\REG[i]$, there is  a sequence of 
values $H_i$ such that, if $p_i$ is correct, $H_i$ is the sequence of values 
written by $p_i$. 
\end{lemma}

\begin{proofL}
Let us define $H_i$ as follows. Let us consider all  the messages
{\sc write}$(-,sn)$ r-delivered from a (correct or faulty) process $p_i$ 
by the correct processes (due to Lemma~\ref{lemma:r-delivery}, these messages 
are r-delivered to all correct processes). 
Let us order these messages according to their processing order as defined 
by the predicate of line~\ref{RB-SWMR-H12}. 
$H_i$ is the corresponding sequence of values. 
(Let us notice that, if $p_i$ is Byzantine, it is possible that some 
of its messages {\sc write}$()$ are r-delivered but never processed
at line~\ref{RB-SWMR-H14}; if any, such messages are never added to $H_i$). 

Let us now  consider the case where $p_i$ is correct. It follows from 
the RB-Validity property of the r-broadcast abstraction that 
any message r-delivered from $p_i$, was r-broadcast by $p_i$. 
It then follows from lines~\ref{RB-SWMR-H01}-\ref{RB-SWMR-H02}
that $H_i$ is the sequence of values written by $p_i$.   
\renewcommand{\toto}{lemma:history}
\end{proofL} 

\Xomit{
\paragraph{Simplifying the notations}
 Let us consider  the sequence $H_k$ associated with the register 
$\REG[k]$ of a  Byzantine process $p_k$. $H_k= v_1,v_2, \cdots$. 
According to the algorithm and the behavior of $p_k$, it is possible 
that the sequence number associated with $v_z$ is not the the integer $z$
(as for the correct processes), but an arbitrary integer 
(line~\ref{RB-SWMR-H01}). Let us observe that the correct processes 
process only the {\sc write}$()$ messages from $p_k$ in increasing order
(line~\ref{RB-SWMR-H12}).

To simplify the presentation, we consider in the following 
that the sequence numbers associated with the values written in any 
register $\REG[i]$ are the consecutive integers $1,2,\cdots$.\footnote{We 
could use a one-to-one correspondence table associating  each position $x$
in $H_k$  with the sequence number $wsn$ appearing in the message
{\sc write}$(v,wsn)$, where $H_k[x]=v$. 
This would uselessly overload the presentation.}  
}

\begin{lemma}
\label{lemma:read-followed-by-write}
Let $p_i$ be a correct process. 
If  $read[i,j,x]$ terminates before  $write[j,y]$ starts, we have  $x<y$.
\end{lemma}

\begin{proofL}
Let $p_i$ a correct process that returns value $v$ from the invocation 
of $\REG[j].{\sf read}()$.  Let $reg_i[j]=\langle v,x\rangle$ the 
pair obtained by $p_i$ at line~\ref{RB-SWMR-H08}, i.e., $v=H_j[x]$ and 
$reg_i[j].sn \geq x$ when $read[i,j,x]$ terminates.

As $write[j,y]$ defines $H_j[y]$, it follows that a message {\sc write}$(-,y)$ 
is r-delivered from $p_j$ at each correct process $p_k$ which executes 
$reg_k[j]\leftarrow \langle -,y\rangle$ at line~\ref{RB-SWMR-H13}. 
As this occurs after  $read[i,j,x]$ has terminated, we necessarily have 
$x<y$. 
\renewcommand{\toto}{lemma:read-followed-by-write}
\end{proofL} 

\begin{lemma}
\label{lemma:write-followed-by-read}
Let  $p_i$ and $p_j$  be  correct processes. 
If $write[i,x]$ terminates before  $read[j,i,y]$ starts, we have  $x\leq y$.
\end{lemma}

\begin{proofL}
Let $p_i$ a correct process that returns from its $x^{\mbox{\small th}}$
invocation of $\REG[i].{\sf write}()$. 
It follows from line~\ref{RB-SWMR-H01}
that  the sequence number $x$ is associated with the written value. 
It follows from the r-broadcast of the message {\sc write}$(v,x)$  
issued  by $p_i$ (line~\ref{RB-SWMR-H02}), and its r-delivery  
(line~\ref{RB-SWMR-H12}) at each correct process (RB-uniformity of the
r-broadcast), that $p_i$ receives  $(n-t)$  messages 
{\sc write\_done}$(x)$ (line~\ref{RB-SWMR-H03}). 
Let $Q_1$ be this set of $(n-t)$ processes that sent these  messages
(line~\ref{RB-SWMR-H14}).
Let us notice that there are at least $(n-2t)$ correct processes in $Q_1$ 
and, due to line~\ref{RB-SWMR-H13}, any of them, say $p_k$, is such that
$reg_k[i].sn\geq x$.

Let $p_j$ be a correct process that invokes  $\REG[i].{\sf read}()$.
The freshness predicate of line~\ref{RB-SWMR-H07} blocks $p_j$ until 
$reg_j[i].sn \geq \max(wsn_1,...,wsn_{n-t})$. 
Let $Q_2$ be the set of the $(n-t)$  processes that sent the
messages {\sc state}$()$ (line~\ref{RB-SWMR-H15}) which allowed 
$p_j$ to exit the wait statement  of line~\ref{RB-SWMR-H07}.   

It follows from Lemma~\ref{lemma:quorum-intersection} that at least one 
correct process $p_k$ belongs to $Q_1\cap Q_2$. 
Hence, when $p_i$ returns from $\REG[i].{\sf write}()$
it received the message {\sc write\_done}$(x)$ from $p_k$, 
and we have then $reg_k[i].sn\geq x$. 
As $\REG[i].{\sf read}()$ by $p_j$ started after  $\REG[i].{\sf write}()$ 
by $p_i$ terminated,  when $p_k$ sends to $p_j$ the message 
{\sc state}$(-,reg_k[i].sn)$, we have $reg_k[i].sn\geq x$. 
It follows that, when $p_j$ exits the wait statement of line~\ref{RB-SWMR-H08} 
we have $reg_j[i].sn \geq x$, which concludes the proof of the lemma.
\renewcommand{\toto}{lemma:write-followed-by-read}
\end{proofL} 

\begin{lemma}
\label{lemma:read-followed-by-read}
Let $p_i$ and $p_j$ be two correct processes. 
If  $read[i,k,x]$ terminates before  $read[j,k,y]$ starts, we have  $x\leq y$.
\end{lemma}

\begin{proofL}
Let us consider process $p_i$. 
When it terminates  $read[i,k,x]$, if follows from the messages 
{\sc catch\_up}$()$ and {\sc catch\_up\_done}$()$ 
(lines~\ref{RB-SWMR-H09}-\ref{RB-SWMR-H10} and
lines~\ref{RB-SWMR-H16}-\ref{RB-SWMR-H17}) that  $p_i$ 
received the acknowledgment message {\sc catch\_up\_done}$(k,x)$ from 
$(n-t)$ different processes. Let $Q_1$ be this set of $(n-t)$ processes.
Let us notice that there are at least $(n-2t)$ correct processes in $Q_1$, 
and for each of them, say  $p_\ell$,  we have  $reg_\ell[k].sn\geq x$.

When $p_j$ invokes $\REG[k].{\sf read}()$ it broadcasts  
the message {\sc read}$()$ and waits until the freshness predicate 
is satisfied (lines~\ref{RB-SWMR-H07}). The messages 
{\sc state}$(-,-)$ it receives are from $(n-t)$ different processes. 
Let $Q_2$ be this set of $(n-t)$  processes.

It follows from Lemma~\ref{lemma:quorum-intersection} that at least one 
correct process $p_\ell$ belongs to $Q_1\cap Q_2$. 
According to the fact that $read[i,k,x]$ terminates before  
$read[j,k,y]$ starts, it follows that $p_\ell$
sent {\sc catch\_up\_done}$(k,x)$ to $p_i$ before 
sending the message {\sc state}$(-,s)$ to $p_j$. 
As $reg_\ell[k].sn$ never decreases, it follows that $x\leq s$. 
It finally follows that, when the freshness predicate is satisfied
at $p_j$, we have $reg_j[k].sn  \geq s$.  As $y=reg_j[k].sn$
(lines~\ref{RB-SWMR-H08}-\ref{RB-SWMR-H11}), it follows that $x\leq y$, 
which concludes the proof.  
\renewcommand{\toto}{lemma:read-followed-by-read}
\end{proofL} 

\subsection{Piecing together the lemmas} 
\begin{theorem}
\label{theorem:atomic-SWMR}
The algorithm 
described in  Figure~{\em\ref{algo:RB-SN-based-SWMR-atomic-register}} 
implements an array of $n$ {\em SWMR} atomic (linearizable) registers 
(one register per process) in the system model ${\cal BAMP}_{n,t}[t<n/3]$.
\end{theorem}

\begin{proofT}
The proof follows from 
Lemmas~\ref{lemma:write-termination}-\ref{lemma:read-followed-by-read}
and  Theorem~\ref{theorem:linearizable}.
\renewcommand{\toto}{theorem:atomic-SWMR}
\end{proofT}

\section{Conclusion}
\label{sec:conclusion} 
This paper presented a signature-free algorithm building an array of $n$ 
single-writer/multi-reader atomic registers (with a register per process) 
in an $n$-process asynchronous message-passing system where up to $t<n/3$ 
processes may commit Byzantine failures.  

This algorithm relies on an underlying reliable broadcast~\cite{B87}, 
an appropriate freshness predicate and a value resynchronization mechanism 
which ensure that a correct process always reads up-to-date values.
A noteworthy property of this algorithm lies in its conceptual simplicity. 

According to the result of~\cite{IRRS14} this algorithm is optimal from
a $t$-resilience point of view. While the cost of a read operation is linear
with respect to $n$, a  problem which remains open lies in its 
$O(n^2)$ message complexity for write operations. This cost is due to 
the use of a Byzantine-tolerant reliable broadcast. Hence the question: 
Is it possible to reduce it, or is $O(n^2)$ a lower bound when one has to 
implement an atomic register in a signature-free message-passing distributed 
system prone to Byzantine failures? We conjecture it is a lower bound. 

\section*{Acknowledgments}
This work has been partially supported by the French ANR project DISPLEXITY
devoted to  computability and complexity in distributed computing,  
the Franco-German ANR-DFG project  DISCMAT (devoted to connections between 
mathematics and distributed computing), and the French ANR project CO$_2$Dim. 



\appendix
\section{A Reliable Broadcast Algorithm}
\label{sec:algorithm-RB}

The r-broadcast algorithm presented in Figure~\ref{algo:reliable-broadcast} 
is Bracha's algorithm~\cite{B87} enriched with sequence numbers. 
Each process $p_i$ manages a local array $next_i[1..n]$, where $next_i[j]$ 
is the sequence number $sn$ of the next application message  (namely, 
{\sc app}$(-,sn)$)  from $p_j$, that $p_i$ will process (line~\ref{G-RB03}).  
Initially, for all $i,j$, $next_i[j]=1$.

\begin{figure}[]
\centering{\fbox{
\begin{minipage}[t]{150mm}
\footnotesize
\renewcommand{\baselinestretch}{2.5}
\resetline
\begin{tabbing}
aaaaa\=aaa\=aaaaa\=aaaaaa\=\kill

{\bf operation} $\mathsf{R\_broadcast}$ {\sc app}($v,sn$):\\
\line{G-RB01}
\> $\mathsf{broadcast}$ {\sc app}$(v,sn)$.\\
~\\
{\bf when a message} {\sc app}$(v,sn)$ {\bf from} $p_j$ 
                               {\bf is} $\mathsf{received}$:\\

\line{G-RB02} 
\> {\bf if} no message {\sc app}$(-,sn))$ $\mathsf{received}$ from $p_j$\\

\line{G-RB03} 
\> \> {\bf then} \= {\bf wait} $(next_i[j]=sn)$;\\

\line{G-RB04} 
\>\> \>  $\mathsf{broadcast}$ {\sc echo}$(j,v,sn)$ \\

\line{G-RB05} 
\>   {\bf end if}.\\~\\

{\bf when a message} {\sc echo}$(j,v,sn)$ {\bf is} $\mathsf{received}$:\\
\line{G-RB06} 
\> {\bf if} \= $~~~$ \= 
   {\sc echo}$(j,v,sn)$ $\mathsf{received}$ 
         from strictly more than $\frac{n+t}{2}$ different processes\\

\line{G-RB07} 
\>          \> $\land$ \> {\sc ready}$(j,v,sn)$ never sent\\

\line{G-RB08} 
\>          \>{\bf then} \=
    $\mathsf{broadcast}$  {\sc ready}$(j,v,sn)$\\

\line{G-RB09} 
\> {\bf end if}.\\
~\\
{\bf when a message} {\sc ready}$(j,v,sn)$ {\bf is} $\mathsf{received}$:\\
\line{G-RB10} 
\> {\bf if} \= $~~~$ \= 
             {\sc ready}$(j,v,sn)$ $\mathsf{received}$ 
                    from at least $t+1$ different processes\\
                    
\line{G-RB11} 
\>          \> $\land$  \>  {\sc ready}$(j,v,sn)$ never  sent\\

\line{G-RB12} 
\>          \>{\bf then} \= $\mathsf{broadcast}$ {\sc ready}$(j,v,sn)$\\

\line{G-RB13} 
\> {\bf end if};\\

\line{G-RB14} 
\> {\bf if} \= $~~~$ \= 
         {\sc ready}$(j,v,sn)$  $\mathsf{received}$ 
                    from at least $2t+1$ different processes\\
                    
\line{G-RB15} 
\>          \> $\land$  \> 
     {\sc app}$(v,sn)$  by $p_j$ never $\mathsf{R\_delivered}$\\

\line{G-RB16}
\>          \>{\bf then} \=
         $\mathsf{R\_deliver}$ {\sc app}$(v,sn)$ from $p_j$;\\

\line{G-RB17}
\>          \>\>    $next_i[j]\leftarrow  next_i[j] +1$\\

\line{G-RB18}
\> {\bf end if}.
\end{tabbing}
\end{minipage}
}
\caption{Reliable Broadcast in ${\cal BAMP}_{n,t}[t<n/3]$,  
(code for process $p_i$)}
\label{algo:reliable-broadcast}
}
\end{figure}

When a process $p_i$ invokes  $\mathsf{R\_broadcast}$ {\sc app}($v,sn$), 
it  broadcasts the message {\sc app}$(v,sn)$ (line~\ref{G-RB01})
where $sn$ is its  next sequence number. 
On its ``server'' role, the behavior of a process $p_i$ is as follows.
\begin{itemize}
\vspace{-0.2cm}
\item
When a process $p_i$ receives a message {\sc app}$(v,sn)$ from a process 
$p_j$ for the first time, it first waits until it can  process this message 
(line~\ref{G-RB03}). Process $p_i$ then  broadcasts a message 
{\sc echo}$(j,v,sn)$  (line~\ref{G-RB04}). If  the message  
just received is not the first  message {\sc app}$(-,sn)$, 
$p_j$ is Byzantine and the message  is discarded. 
\vspace{-0.2cm}
\item
Then, when  $p_i$ has received the  same message {\sc echo}$(j,v,sn)$  
from ``enough'' processes (where ``enough'' means 
``more than $(n+t)/2$ different processes''), and  has not yet broadcast a 
message {\sc ready}$(j,v,sn)$, it does it (lines~\ref{G-RB06}-\ref{G-RB09}). 

The aim of (a) the messages {\sc echo}$(j,v,sn)$,  and (b) the cardinality 
``greater than  $(n+t)/2$ processes'', is to ensure that no two correct 
processes can r-deliver distinct messages from $p_j$ (in the case where $p_j$ 
is Byzantine). 
The  aim of the messages {\sc ready}$(j,v,sn)$  is related to the liveness 
of the  algorithm. Namely,   its aim is  to allow  (at least when  $p_j$ is
correct)  the  r-delivery by the correct processes  of the very 
same message {\sc app}$(v,sn)$ from $p_j$, and this must always occur if 
$p_j$ is correct. It is nevertheless possible that a message r-broadcast 
by a  Byzantine process $p_j$ be never r-delivered by the correct processes. 
\vspace{-0.2cm}
\item
Finally, when $p_i$  has received the  message {\sc ready}$(j,v,sn)$
from $(t+1)$ different  processes, it broadcasts the same  message 
{\sc ready}$(j,v,sn)$, it not yet done. This is required to ensure 
the RB-termination property. If $p_i$ has received ``enough'' 
 messages {\sc ready}$(j,v,sn)$ (as before ``enough'' means
``from more than $(n+t)/2$ different processes''), it 
r-delivers the  message {\sc app}$(v,sn)$ r-broadcast by $p_j$. 
\end{itemize}

Proofs that this algorithm satisfies the properties defining 
the reliable broadcast abstraction can be found in~\cite{B87,MR14}.

\end{document}